\documentstyle[aps,psfig]{revtex}

\begin{document}

\title{A generic estimate of trans-Planckian modifications to the primordial power spectrum in inflation}
\author{Richard Easther$^1$, Brian R. Greene$^{1,2}$, William
H. Kinney$^1$, Gary Shiu$^3$
}
\address{$^1$ Institute for Strings, Cosmology and Astroparticle
  Physics, Columbia University, New York, NY 10027, USA}
\address{$^2$ Department of Mathematics, Columbia University, New York,
NY 10027,   USA}
\address{$^3$ Department of Physics and Astronomy, University
  of Pennsylvania,
Philadelphia, PA 19104, USA
\\ \mbox{}\\Email:     easther@phys.columbia.edu, greene@phys.columbia.edu,
kinney@phys.columbia.edu, shiu@dept.physics.upenn.edu}
\address{}

\date{April 16, 2002}
\maketitle

\begin{abstract}
We derive a general expression for the power spectra of scalar and tensor fluctuations generated during inflation given an arbitrary choice of boundary condition for the mode function at a short distance. We assume that the boundary condition is specified at a short-distance cutoff at a scale $M$ which is independent of time. Using a particular prescription for the boundary condition at momentum $p \sim M$, we find that the modulation to the power spectra of density and gravitational wave fluctuations is of order $(H/M)$, where $H$ is the Hubble parameter during inflation, and we argue that this behavior is generic, although by no means inevitable. With fixed boundary condition, we find that the shape of the modulation to the power spectra is determined entirely by the deviation of the background spacetime from the de Sitter limit.
\end{abstract}


\baselineskip=20pt

\section{Introduction}
There is currently significant interest in the possibility that Planck
scale physics might leave an observable signature in the cosmic
microwave background radiation. The common thread running through this
work is that early universe quantum field fluctuations
may be sensitive to short distance physics, and this sensitivity may
induce small modifications to the CMB (Cosmic Microwave Background)
spectrum produced by inflation. A variety of approaches have been taken to
calculate the precise signature and a variety of results have been
found. These include modifications to the dispersion relation for
quantum modes at short distance, $\omega^2 \neq p^2 +
m^2$\cite{brandenberger00a,martin00,niemeyer00,niemeyer01},
string-inspired changes to the Heisenberg uncertainty
relation\cite{kempf00,kempf01,easther00,easther00a,hassan02}, and 
noncommutative geometry\cite{CGS,lizz02,brandenberger02}.

In this paper, we look at the problem of modifications to the power spectrum from short-distance effects in a generic way, independent of the specific nature of the short-distance physics assumed. We make the supposition that the modification to physics at short distances has no explicit time dependence. That is, there will be a cutoff momentum $p_{\rm c}$ which is given by a physical constant such as the Planck mass or the string scale that is independent of time. Any change in standard physics at short distance or high momentum $p > p_{\rm c}$ will be completely described (in a phenomenological sense) by the boundary condition imposed on quantum modes at the cutoff scale. The boundary conditions at the cutoff are equivalent to the selection of a vacuum state for the theory. In this paper, we derive a general expression for the change in the scalar and tensor power spectra which results from an arbitrary choice of vacuum at short distance.

We apply this result to the ansatz of Danielsson\cite{danielsson02} for selecting a boundary condition for the mode function. Danielsson's proposal is simply that modes are ``created'' at the cutoff length in a local Minkowski state, even though the vacuum state of the spacetime is slightly rotated from the Minkowski vacuum. We generalize Danielsson's result to an arbitrary non-de Sitter background.  This ansatz clearly falls short of being a complete description of physics at short distances. However, it does represent a phenomenologically reasonable way to estimate the magnitude and form of the changes to the primordial power spectrum due to the presence of a short-distance cutoff. For a cutoff momentum $p_{\rm c} \sim M$, Kaloper et al.\cite{shenker01} have argued that the largest effect consistent with effective field theory was of order $(H / M)^2$, where $H << M$ is the Hubble constant during inflation. This differed from the conclusion of Easther, et al.\cite{easther00,easther00a} that the effect in a particular well-motivated model\cite{kempf00} was of order $H/M$, potentially large enough to have an observable effect on the primordial power spectrum. Brandenberger and Martin\cite{brandenberger01} and subsequently Danielsson\cite{danielsson02} have argued that deviations of the quantum state of the field from the local vacuum state generically result in an effect of order $H/M$, and we adopt this point of view here. Certainly the final resolution of this issue awaits the formulation of a more complete theory of physics at short distances. In the meantime, a phenomenological perspective can potentially allow us to constrain the form of such a theory using astrophysical observations, which is a tantalizing prospect.

The paper is structured as follows: Section II contains the calculation of the scalar and tensor power spectra for a general choice of vacuum. Section III considers the case of the Danielsson ansatz for an adiabatic vacuum. Section IV contains a summary and conclusions. A review of the generation of perturbations in inflation, including definitions of some parameters used in the paper, are contained in an Appendix.

\section{The mode function for arbitrary vacuum}

It can be shown that the scalar fluctuation modes $u_k$ and tensor fluctuation modes $v_k$ generated during inflation obey the following equations of motion:
\begin{equation}
y^2 \left(1 - \epsilon\right)^2 {d^2 u_k \over d y^2} + 2 y \epsilon \left(\epsilon - \eta\right) {d u_k \over d y} + \left[y^2 - F\left(\epsilon, \eta, \xi\right)\right] u_k = 0,\label{eqexactmodeequationy}
\end{equation}
where 
\begin{equation}
F\left(\epsilon,\eta,\xi\right) \equiv 2 \left(1 + \epsilon - {3 \over 2} \eta + \epsilon^2 - 2 \epsilon \eta + {1 \over 2} \eta^2 + {1 \over 2} \xi^2\right),
\end{equation}
and
\begin{equation}
y^2 \left(1 - \epsilon\right)^2 {d^2 v_k \over d y^2} + 2 y \epsilon \left(\epsilon - \eta\right) {d v_k \over d y} + \left[y^2 - \left(2 - \epsilon\right)\right] v_k = 0.\label{eqtenexactmodeequationy}
\end{equation}
Here $\epsilon, \eta, \xi$ are slow-roll parameters (see Appendix), and the ``time'' variable $y$ is defined to be the wavelength of the mode relative to the horizon size:
\begin{equation}
y \equiv {k \over a H}.
\end{equation}
The variable $y$ can be related to the conformal time $\tau$ by:
\begin{equation}
dy = -k \left(1 - \epsilon\right) d\tau,
\end{equation}
where during an inflationary phase the conformal time is negative, $\tau < 0$, with $d\tau > 0$ and $\epsilon < 1$. We emphasize that these equations are exact in the sense that no assumption of slow roll has been made.

We will assume that at wavelengths much longer than some cutoff distance, $\lambda \gg \lambda_C$, the mode equations (\ref{eqexactmodeequationy}) and (\ref{eqtenexactmodeequationy}) describe the evolution of the scalar and tensor mode functions, respectively. For definiteness, we begin with the power-law inflation case and then extend the calculation to the slow-roll case. For power-law inflation, $\epsilon = \eta = \xi = {\rm const.}$, and the scale factor obeys $a \propto t^{1/\epsilon}$. In this case, the mode equations for scalars and tensors become identical:
\begin{equation}
y^2 \left(1 - \epsilon\right)^2 {d^2 u_k \over d y^2} + \left[y^2 - (2 - \epsilon)\right] u_k = 0\label{eqpowerlawmodey}
\end{equation}
for scalar fluctuations and 
\begin{equation}
y^2 \left(1 - \epsilon\right)^2 {d^2 v_k \over d y^2} + \left[y^2 - (2 - \epsilon)\right] v_k = 0\label{eqtenpowerlawmodey}
\end{equation}
for tensor fluctuations. We will confine our discussion of the mode behavior to the scalar fluctuations with the understanding that the properties of the solutions apply equally well to tensor modes. Since $\epsilon = {\rm const.}$, the exact solution to the mode equations is
\begin{equation}
u_k = {1 \over 2} \sqrt{\pi \over k} \sqrt{y \over 1 - \epsilon}\left[C_{+} H_\nu\left(y \over 1 - \epsilon\right) + C_{-} H^*_\nu\left(y \over 1 - \epsilon\right)\right].\label{eqpowerlawmodefunc}
\end{equation}
Here the variable $y$ is just the wavelength of the mode relative to the horizon size, $y \equiv k / \left(a H\right)$, and 
\begin{equation}
\nu = {3 - \epsilon \over 2 \left(1 - \epsilon\right)}.\label{eqpowerlawnu}
\end{equation}
Note in particular that this solution is equally valid for modes inside the horizon and modes outside the horizon. If the solution is fixed by a boundary condition well inside the horizon, $y = k / \left(a H\right) \gg 1$, this solution can be seamlessly followed to the long-wavelength limit, $y \rightarrow 0$ without the need for ``matching'' solutions at horizon crossing ($y = 1$). The case of a Bunch-Davies vacuum is then exactly equivalent to selecting $C_{+} = 1$ and $C_{-} = 0$ in the general solution (\ref{eqpowerlawmodefunc}). The canonical commutation relation for the field requires
$\left|C_{+}\right|^2 - \left|C_{-}\right|^2 = 1$.
Note that since the fluctuation amplitude $P_{\cal R} \propto \left|u_k\right|$ depends only on the amplitude of the mode function, the problem contains an arbitrary overall phase.  

Previous investigations of whether an observable signature of short-distance physics in inflation is possible\cite{brandenberger00a,martin00,niemeyer00,niemeyer01,kempf00,kempf01,easther00,easther00a,lizz02,brandenberger02} have approached the problem by defining a particular model of short-distance physics and then evaluating the influence of that ansatz on the amplitude of the modes at long wavelength. Here we use a more generic approach: {\em whatever} the nature of the short-distance physics, if we assume that the mode equations (\ref{eqexactmodeequationy}) and (\ref{eqtenexactmodeequationy}) are valid at long wavelength, then the only influence the short-distance physics can have on the behavior of the modes at long wavelength is through the choice of the coefficients $C_{\pm}$. (This fact was noted by Starobinsky\cite{starobinsky01}.) However, the coefficients $C_{\pm}$ in general need not be constant: they may also have an intrinsic $k$-dependence. An important question is: can we determine the $k$-dependence of the coefficients without knowing the precise nature of the short-distance physics? In this section we show that the dependence of $C_{\pm}\left(k\right)$ on $k$ depends only on the nature of the background spacetime, as long as there is no explicit time-dependence in the boundary condition at short wavelength. (This condition will be made precise below.)

We wish to introduce a modification to standard quantum field theory at short distances. That is, for states with momentum higher than some cutoff $p > p_{\rm c}$, the standard equations of motion for the modes no longer apply. Brandenberger and Martin\cite{brandenberger00a,martin00} and Niemeyer\cite{niemeyer00} proposed modifying the dispersion relation of the modes at high momentum $\omega^2 \neq p^2 + m^2$. Kempf proposed a cutoff based on a modification to the uncertainty relation\cite{kempf00,kempf01} $\Delta x \Delta p \geq \left(1 / 2\right) \left[1 + \beta \left(\Delta p\right)^2 + \cdots\right]$. Other choices are certainly possible\cite{lizz02,brandenberger02}. However, whatever choice we wish to make for the nature of the physics at momentum $p > p_{\rm c}$, once a given state is redshifted to low momentum $p \ll p_{\rm c}$, the standard mode equation (\ref{eqpowerlawmodey}) holds and the solutions will be of the form (\ref{eqpowerlawmodefunc}). Thus, any choice of short-distance physics can be mapped to a choice of the coefficients $C_{\pm}\left(k\right)$. We will therefore consider an arbitrary boundary condition fixing the $C_{\pm}$ at the cutoff momentum $p_{\rm c}$:
\begin{eqnarray}
C_{\pm} = C_{\pm}\left(k/a = p_{\rm c}\right).
\end{eqnarray}
We will make the assumption that this boundary condition contains no explicit time dependence: that is, the cutoff momentum $p_{\rm c}$ is constant in time. Taking the cutoff to be at a scale much smaller than the horizon, $p_c \gg H$, the mode function becomes
\begin{eqnarray}
u_k \rightarrow C_{\rm +} e^{i y / \left(1 - \epsilon\right)} + C_{\rm -} e^{-i y / \left(1 - \epsilon\right)},\ y \gg 1.\label{eqasymptoticmodefunc}
\end{eqnarray}
Since at short wavelength the mode function $u_k$ depends only on $y = k / \left(a H\right) = p / H$, the boundary condition on the mode function is set by specifying the values of $u_k\left(y\right)$ and its derivative at $ y = y_{\rm c} \equiv p_{\rm c} / H$:
\begin{eqnarray}
C_{\rm +} =&& {1 \over 2} e^{-i y_{\rm c} / \left(1 - \epsilon\right)} \left[u_k\left(y_c\right) - i \left(1 - \epsilon\right) u_k'\left(y_c\right)\right]\cr
C_{\rm -} =&& {1 \over 2} e^{+i y_{\rm c} / \left(1 - \epsilon\right)} \left[u_k\left(y_c\right) + i \left(1 - \epsilon\right) u_k'\left(y_c\right)\right].\label{eqCpmbc}
\end{eqnarray}
Note that the $k$-dependence of the coefficients is then entirely contained in the  boundary value $y_c$. This is due to the dependence of $y_{\rm c}$ on the Hubble parameter, which is in general time dependent:
\begin{equation}
y_{\rm c}\left(k\right) \equiv {k \over a H}\bigg|_{\left(k / a\right) = p_{\rm c}}.
\end{equation}
That is, $y_{\rm c}\left(k\right)$ is defined to be the value of $k / \left(a H\right)$ at the time when a given comoving mode $k$ is at the cutoff scale $\left(k / a\right) = p_{\rm c}$ in physical units. Especially important is that in the limit of de Sitter space, $H = {\rm const.}$ and therefore $y_c = {\rm const.}$, so that the coefficients $C_{\pm}$ are generically independent of $k$ in the de Sitter limit. The power spectrum $P_{\cal R}$ is defined in the long-wavelength limit $y \rightarrow 0$, where 
\begin{equation}
\left|u_{k}\right| \rightarrow {2^{\nu - 3/2} \over \sqrt{2 k}} {\Gamma\left(\nu\right) \over \Gamma\left(3/2\right)} \left|C_{+}\left(k\right) + C_{-}\left(k\right)\right| \left({y \over 1 - \epsilon}\right)^{1/2 - \nu}.
\end{equation}
The scalar power spectrum is then
\begin{equation}
P_{\cal R}^{1/2} = \sqrt{k^3 \over 2 \pi^2} \left|{u_k \over z}\right| = {2^{\nu - 3/2} \over 2 \pi} {\Gamma\left(\nu\right) \over \Gamma\left(3 /2\right)} \left({H \over m_{\rm Pl} \sqrt{\epsilon}}\right) \left|C_{\rm+}\left(k\right) + C_{-}\left(k\right)\right| \left({y \over 1 - \epsilon}\right)^{3/2 - \nu},\label{eqscalarpowerspectrum}
\end{equation}
where the $C_{\pm}$ are given by Eq. (\ref{eqCpmbc}), and in the case of power-law inflation, $\nu$ is given by Eq. (\ref{eqpowerlawnu}). This is trivially generalized to the case of slow-roll inflation, which assumes $\epsilon$ and $\eta$ are constant and small but unequal, so that
\begin{equation}
F\left(\epsilon,\eta,\xi\right) \simeq 2 \left(1 + \epsilon - {3 \over 2} \eta \right) \simeq {\rm const.}
\end{equation}
The solutions to the mode equation are then of the form (\ref{eqscalarpowerspectrum}), but with 
\begin{equation}
\nu = {3 \over 2} + 2 \epsilon - \eta.
\end{equation}
Note that for the power-law case, $\epsilon = \eta$, this is simply the small $\epsilon$ limit of Eq. (\ref{eqpowerlawnu}). Similarly, the tensor fluctuation amplitude is
\begin{equation}
P_{\cal T}^{1/2}\left(k\right) = \sqrt{k^3 \over 2 \pi^2} \left|{v_k \over m_{\rm Pl} a}\right| = 
{2^{\nu - 3/2} \over 2 \pi} {\Gamma\left(\nu\right) \over \Gamma\left(3 /2\right)}\left({H \over m_{\rm Pl}}\right) \left|C_{\rm+}\left(k\right) + C_{-}\left(k\right)\right| \left({y \over 1 - \epsilon}\right)^{3/2 - \nu},
\end{equation}
where
\begin{equation}
\nu = {3 - \epsilon \over 2 \left(1 - \epsilon\right)}
\end{equation}
in the power law case, and 
\begin{equation}
\nu = {3 \over 2} - \epsilon
\end{equation}
in the slow roll case. We assume that the boundary conditions are identical for the scalar and the tensor modes, and thus that the $C_{\pm}\left(k\right)$ are identical for scalars and tensors. In the case of Bunch-Davies vacuum, $C_{+} = 1$ and $C_{-} = 0$, the spectra $P_{\cal R}$ and $P_{\cal T}$ are power-law in $k$, and we can calculate the spectral indices:
\begin{eqnarray}
n_{\cal R} \equiv&& 1 + {d \log{P_{\cal R}} \over d \log{k}}\cr
=&& 1 + {d \log{P_{\cal R}} \over d \log{y}}\bigg|_{(a H) = {\rm const.}} = 4 - 2 \nu,
\end{eqnarray}
for the scalar modes and 
\begin{eqnarray}
n_{\cal T} \equiv&&  {d \log{P_{\cal T}} \over d \log{k}}\cr
=&& {d \log{P_{\cal T}} \over d \log{y}}\bigg|_{(a H) = {\rm const.}} = 3 - 2 \nu.
\end{eqnarray}
These are the standard results. However, a choice of vacuum different from Bunch-Davies will in general lead to an additional $k$-dependence in the spectrum. One possible observational signature of a modulation of the power spectrum is a violation of the inflationary ``consistency condition'', which relates the ratio of the tensor and scalar amplitudes to the tensor spectral index:
\begin{equation}
{P_{\cal T} \over P_{\cal R}} = - {1 \over 2} n_{\cal T}.
\end{equation} 
Modulation of the power spectra leaves the tensor/scalar ratio unchanged, but alters the spectral index $n_{\cal T}$, leading to a violation of the consistency condition\cite{hui00}. 

To determine the specific form of the change in the power spectra, we must determine $y_{\rm c}\left(k\right)$. This can be done as follows. Take the function $k\left(\tau\right)$ as the wavenumber corresponding to the cutoff scale $p_{\rm c}$ at conformal time $\tau$,
\begin{equation}
k\left(\tau\right) \propto a\left(\tau\right).
\end{equation}
Then we can calculate
\begin{equation}
{d \log{y_{\rm c}} \over d \log{k}} = {d \log{H^{-1}} \over d \log{a}} = {d \log{H^{-1}} \over d \phi} {d \phi \over d \tau} {d \tau \over d\log{a}}.
\end{equation}
Working out the terms individually, we have (see Appendix)
\begin{equation}
{d \log{H^{-1}} \over d \phi} = - {H'\left(\phi\right) \over H\left(\phi\right)},
\end{equation}
\begin{equation}
{d \phi \over d \tau} = a \dot\phi = - {m_{\rm Pl}^2 \over 4 \pi} a H'\left(\phi\right),
\end{equation}
and 
\begin{equation}
{d \tau \over d \log{a}} = {1 \over a H}.
\end{equation}
Then 
\begin{equation}
{d \log{y_{\rm c}} \over d \log{k}} = {m_{\rm Pl}^2 \over 4 \pi} \left[{H'\left(\phi\right) \over H\left(\phi\right)}\right]^2 = \epsilon.
\end{equation}
This equation is exact. Therefore the dependence of $y_c$ on $k$ is particularly simple
\begin{equation}
y_c\left(k\right) \propto k^{\epsilon},
\end{equation}
a result which holds {\em in general}, including the power-law and slow-roll cases. We then have a very simple expression for the modification to the standard spectra arising from a choice of vacuum different from Bunch-Davies:
\begin{equation}
P_{\cal R, T} = \left|C_{\rm +}\left(k\right) + C_{\rm -}\left(k\right)\right|^2 P^{\rm B-D}_{\cal R, T},\label{eqthebigkahuna}
\end{equation}
where the coefficients $C_{\pm}(k)$ are given by Eq. (\ref{eqCpmbc}) and $y_c(k) \propto k^{\epsilon}$. This solution is valid in both the slow roll and power-law cases. The important feature is that the $k$-dependence of the modulation depends entirely on the behavior of the background spacetime, that is the time dependence of $y_c \equiv k / a H$. In particular, in the de Sitter limit $\epsilon \rightarrow 0$, and $y_{\rm c} \rightarrow {\rm const.}$, so that there is no additional $k$-dependence in the power spectrum, although the normalization can be altered. A concrete example of this was studied in \cite{easther00}. In the general (non-de Sitter) case, the boundary value $y_c$ is time-dependent, and we expect a modulation of the power spectrum\cite{easther00a}.

\section{Boundary conditions for the adiabatic vacuum}

Danielsson\cite{danielsson02} has proposed a general ansatz for parameterizing the choice of vacuum in an inflationary spacetime. The choice of Bunch-Davies vacuum can be expressed as a relation between the field and its conjugate momentum in the ultraviolet limit,
\begin{equation}
{d \phi_p \over d t} \longrightarrow - i p \phi_p,\ p \rightarrow \infty.
\end{equation}
Danielsson's prescription for an adiabatic vacuum is to enforce this relation at a {\em finite} momentum $p_{c}$:
\begin{equation}
{d \phi_p \over d t} \equiv - i p \phi_p,\ p = p_c.
\end{equation}
This is simply translated into comoving variables,
\begin{equation}
{1 \over a} {d \over d \tau} \left({u_k \over a}\right) \equiv -i {k \over a^2} u_k,\ (k / a) = (k / a)_c.
\end{equation}
In terms of the variable $y$, the Danielsson boundary condition becomes
\begin{equation}
u_k'\left(y_c\right) = \left[{i y_c - 1 \over y_c \left(1 - \epsilon\right)}\right] u_k\left(y_c\right).
\end{equation}
The adiabatic vacuum can be translated into coefficients $C_{\pm}$ using Eq. (\ref{eqCpmbc}), that is using the {\em asymptotic} ($k \rightarrow \infty$) form of the mode function $u_k$\footnote{We note that if we use the exact solution for the mode function (\ref{eqpowerlawmodefunc}) instead of the asymptotic version (\ref{eqasymptoticmodefunc}), the vacuum rotation will necessarily be exactly zero. The essential physics of this ansatz is that we assume the mode is ``created'' in the local Minkowski state, but the true vacuum of the spacetime is slightly non-Minkowski.}:
\begin{eqnarray}
C_{+} &&= {1 \over 2} e^{-i y_c / \left(1 - \epsilon\right)} \left({2 y_c  + i \over y_c}\right) u_k\left(y_c\right)\cr
C_{-} &&=-  {1 \over 2} e^{+i y_c / \left(1 - \epsilon\right)} \left({ i \over y_c }\right) u_k\left(y_c\right),
\end{eqnarray}
Or,
\begin{equation}
C_{-} = - e^{2 i y_c / \left(1 - \epsilon\right)} \left({i \over 2 y_c + i}\right) C_{+}.
\end{equation}
This is a generalization of the result of Danielsson to arbitrary background. Note that this condition completely specifies the modulation to the power spectrum, since the Wronskian condition 
$\left\vert C_{+} \right\vert^2 - \left\vert C_{-} \right\vert^2  = 1$
 constrains $C_{+}$:
\begin{equation}
\left\vert C_{+} \right\vert^2 = {1 + 4 y_c^2  \over 4 y_c^2}.
\end{equation}
In the limit $y_c \rightarrow \infty$, we have $\left\vert C_{+} \right\vert \rightarrow 1$, as expected. We are free to choose the overall phase, so we can write the two coefficients  as
\begin{eqnarray}
C_{+} = {2 y_c + i \over 2 y_c} e^{-i y_c / \left(1 - \epsilon\right)},\cr
C_{-} = - { i \over 2 y_c } e^{+i y_c / \left(1 - \epsilon\right)}.
\end{eqnarray}
With $y_c \propto k^{\epsilon}$, the modulation of the power spectrum is then completely determined. Note also that the amplitude of the vacuum rotation $C_{-}$ is of order $y_c^{-1}$. If we consider a theory with a cutoff at a scale $p_c \sim M$, the modulation of the power spectrum is then first order in $H/M$:
\begin{equation}
C_{-} \sim y_c^{-1} \sim {H \over M}.
\end{equation}
We expect in general that if short-scale physics results in a rotation of the vacuum, then the lowest-order correction to $C_{\pm}$ will be of this form\cite{danielsson02,easther00,easther00a}. From Eq. (\ref{eqthebigkahuna}), the correction to the power spectrum is then also generically of order $H/M$. This is consistent with the modulation found in Refs. \cite{easther00,easther00a,brandenberger01,danielsson02}. However, it must be emphasized that this answer depends on the choice of boundary condition for the mode, and is not forced on us by the physics. Other choices are possible -- including the choice of Bunch-Davies vacuum with $C_{-} = 0$! Only a more complete understanding of short scale physics will fix the boundary conditions from first principles.

Figs. 1 and 2 show the power spectrum and modulation, respectively, for power-law inflation with $\epsilon = 0.01$. We have chosen $y_c \sim (M / H) = 100$, a value consistent with a cutoff determined by the string scale in conventional string theory, which can be as much as two orders of magnitude larger than the Planck length\cite{kaplun92}. The modulation to the power spectrum is of order one percent -- a potentially observable value.
\begin{figure}
\centerline{\psfig{figure=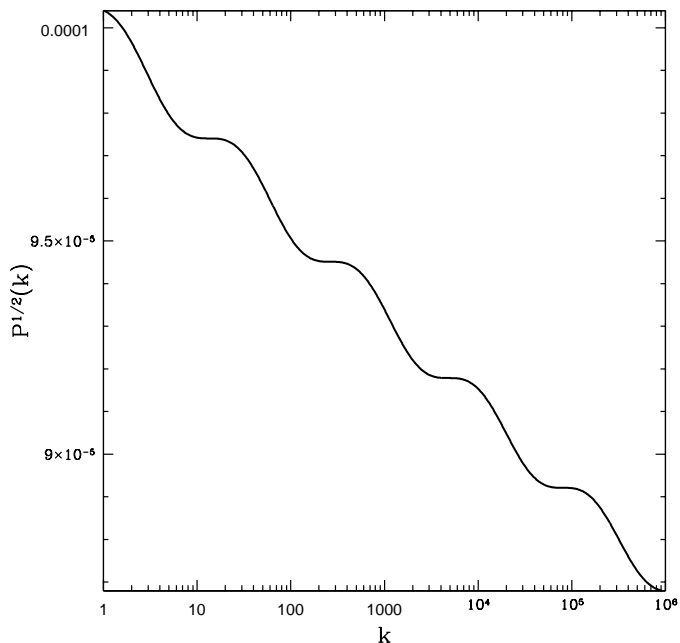,width=3.5in}} \caption{Power spectrum $P^{1/2}(k)$ as a function of $k$, for $\epsilon = 0.01$ and $y_c(k_0) = 100$.}
\end{figure}
\begin{figure}
\centerline{\psfig{figure=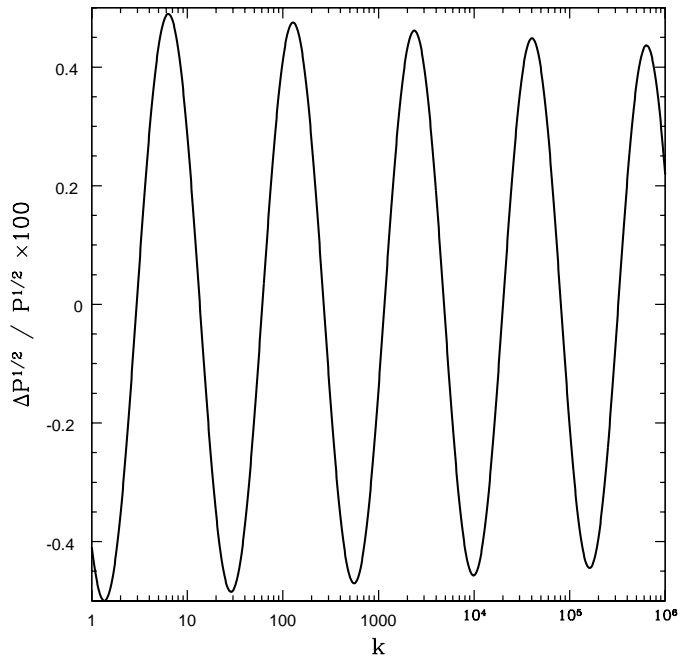,width=3.5in}} \caption{The modulation $\Delta P^{1/2} / P^{1/2} = 1 - \left\vert C_{+} + C_{-} \right\vert$ as a function of wavenumber $k$, for $\epsilon = 0.01$ and $y_c(k_0) = 100$.}
\end{figure}

\section{Conclusions}

We have shown that, since any new physics at short distance will influence the modes generated during inflation through the choice of boundary condition, it is possible to derive a generic formula for the form of the quantum modes at long wavelength. In the limit of de Sitter space, we recover a scale-invariant perturbation spectrum regardless of the choice of boundary condition, although with altered amplitude. However, for non-de Sitter backgrounds, the selection of boundary condition has a potentially large effect on the form of the primordial power spectrum, both for scalar modes and tensor modes.  Allowing freedom in the choice of vacuum opens up a very rich set of possibilities for the primordial power spectrum, since particular instances of the general solution (\ref{eqthebigkahuna}) can in principle take on a very complex form. Tilting the spectra toward the red or the blue is possible, as well as introducing single features or oscillatory behavior, depending on the background and the choice of boundary condition. However, we find in general that the scale-dependence of the modulation is determined by the deviation of the background spacetime from the de Sitter limit and not by the particular form of the boundary conditions.

We study the particular case of the ansatz proposed by Danielsson for selecting the boundary condition for the mode function at the short-distance cutoff\cite{danielsson02}, and extend the result of Ref. \cite{danielsson02} to the case of arbitrary background. We find that deviations from the standard power spectra are generically of order $H/M$, where $H$ is the Hubble constant during inflation, and $M$ is the cutoff scale. This scenario provides a much simpler realization of the effects studied in Refs. \cite{easther00,easther00a} and suggests that the relevant physics determining the amplitude of the modulation is due to the presence of a fixed cutoff scale $M$, and not, for example, due to the breaking of Lorentz invariance. While the amplitude of the modulation depends sensitively on boundary conditions set by unknown short-scale physics, we find it exciting that there are well-motivated choices that yield changes to the primordial power spectrum at the edge of detectability with presently foreseen experiments\cite{easther00a}.

\section*{Acknowledgments}

We would like to thank Lam Hui for many helpful conversations. The
work of BG is supported in part by DOE grant DE-FG02-92ER40699B and
the work of GS was supported in part by the DOE grants
DE-FG02-95ER40893, DE-EY-76-02-3071 and the University of Pennsylvania
School of Arts and Sciences Dean's funds. ISCAP gratefully
acknowledges the generous support of the Ohrstrom Foundation.

\section*{Appendix: Inflation and the production of fluctuations}
In this section we review the basics of scalar field dynamics in inflationary cosmology with emphasis on the very useful Hamilton-Jacobi formalism\cite{grishchuk88,muslimov90,salopek90}. The emphasis is pedagogical; a more formal review can be found in Ref. \cite{lidsey95}. The first basic ingredient is a cosmological metric, which we shall take to be of the flat Robertson-Walker form
\begin{equation}
ds^2 = dt^2 - a^2\left(t\right) \left|d {\bf x}\right|^2 = a^2\left(\tau\right) \left[d\tau^2 - \left|d {\bf x}\right|^2\right].
\end{equation}
The quantity $\tau$ is the conformal time, with $dt = a d\tau$. The second ingredient is a spatially homogeneous scalar field $\phi$ with potential $V\left(\phi\right)$ and equation of motion
\begin{equation}
\ddot \phi + 3 H \dot \phi + V'\left(\phi\right) = 0,\label{eqequationofmotion}
\end{equation}
where the Hubble parameter $H$ is defined as
\begin{equation}
H \equiv \left({\dot a \over a}\right).
\end{equation}
An overdot denotes a derivative with respect to the coordinate time $t$. If the stress energy of the universe is dominated by the scalar field $\phi$, the Einstein field equations for the evolution of the background metric $G_{\mu\nu} = 8 \pi G T_{\mu\nu}$ can be written as
\begin{equation}
H^2 = \left({\dot a \over a}\right)^2 = {8 \pi \over 3 m_{\rm Pl}^2} \left[V\left(\phi\right) + {1 \over 2} \dot\phi^2\right]\label{eqbackgroundequation1}
\end{equation}
and
\begin{equation}
\left({\ddot a \over a}\right) = {8 \pi \over 3 m_{\rm Pl}^2} \left[V\left(\phi\right) - \dot\phi^2\right],\label{eqbackgroundequation2}
\end{equation}
where $m_{\rm Pl} = G^{-1/2} \simeq 10^{19}\,{\rm GeV}$ is the Planck mass. These background equations, along with the equation of motion (\ref{eqequationofmotion}), form a coupled set of differential equations describing the evolution of the universe. The fundamental quantities to be calculated are $\phi\left(t\right)$ and $a\left(t\right)$, and the potential $V\left(\phi\right)$ is input from some model. Inflation is defined to be a period of accelerated expansion 
\begin{equation}
\left(\ddot a \over a\right) > 0,\label{eqconditionforinflation}
\end{equation}
indicating an equation of state in which vacuum energy dominates over the kinetic energy of the field $\dot\phi^2 < V\left(\phi\right)$. In the limit that $\dot\phi = 0$, the expansion of the universe is of the de Sitter form, with the scale factor increasing exponentially in time:
\begin{eqnarray}
&&H = \sqrt{\left({8 \pi \over 3 m_{\rm Pl}^2}\right) V\left(\phi\right)} = {\rm const},\cr
&&a \propto e^{H t}.
\end{eqnarray}
Note that with the Hubble distance $H^{-1}$ constant and the scale factor increasing exponentially, comoving length scales initially smaller than the horizon are rapidly redshifted outside the horizon. In general, the Hubble parameter $H$ will not be exactly constant, but will vary as the field $\phi$ evolves along the potential $V\left(\phi\right)$. A convenient approach to the more general case is to express the Hubble parameter directly as a function of the field $\phi$ instead of as a function of time, $H = H\left(\phi\right)$. This is consistent as long as $t$ is a single-valued function of $\phi$. Differentiating Eq. (\ref{eqbackgroundequation1}) with respect to time,
\begin{eqnarray}
2 H\left(\phi\right) H'\left(\phi\right) \dot\phi =&& \left({8 \pi \over 3 m_{\rm Pl}^2}\right) \left[V'\left(\phi\right) + \ddot\phi\right] \dot \phi\cr
=&& - \left({8 \pi \over m_{\rm Pl}^2}\right) H\left(\phi\right) \dot\phi^2.
\end{eqnarray}
The equation of motion (\ref{eqequationofmotion}) was used to simplify the right-hand side. Substituting back into the definition of $H$ in Eq. (\ref{eqbackgroundequation1}) results in the system of two first-order equations
\begin{eqnarray}
&&\dot\phi = -{m_{\rm Pl}^2 \over 4 \pi} H'\left(\phi\right),\cr
&&\left[H'\left(\phi\right)\right]^2 - {12 \pi \over m_{\rm Pl}^2} H^2\left(\phi\right) = - {32 \pi^2 \over m_{\rm Pl}^4} V\left(\phi\right).\label{eqbasichjequations}
\end{eqnarray}
These equations are completely equivalent to the second-order equation of motion ({\ref{eqequationofmotion}). The second of these is referred to as the {\it Hamilton-Jacobi} equation, and can be written in the useful form
\begin{equation}
H^2\left(\phi\right) \left[1 - {1\over 3} \epsilon\left(\phi\right)\right] = \left({8 \pi  \over 3 m_{\rm Pl}^2}\right) V\left(\phi\right),\label{eqhubblehamiltonjacobi}
\end{equation}
where the parameter $\epsilon$ is defined as
\begin{equation}
\epsilon \equiv {m_{\rm Pl}^2 \over 4 \pi} \left({H'\left(\phi\right) \over H\left(\phi\right)}\right)^2.\label{eqdefofepsilon}
\end{equation}
The physical meaning of the parameter $\epsilon$ can be seen by expressing Eq.  (\ref{eqbackgroundequation2}) as
\begin{equation}
\left({\ddot a \over a}\right) = H^2 \left(\phi\right) \left[1 - \epsilon\left(\phi\right)\right],
\end{equation}
so that the condition for inflation (\ref{eqconditionforinflation}) is given simply by $\epsilon < 1$. Equivalently, $\epsilon$ can be viewed as parameterizing the equation of state of the scalar field matter, with the pressure $p$ and energy density $\rho$ related as
\begin{equation}
p = \rho \left({2 \over 3} \epsilon - 1\right).\label{eqeqnofstate}
\end{equation}
The condition for inflation $\epsilon < 1$ is the same as $\rho + 3 p < 0$.
The de Sitter case is $\epsilon = 0$ or $p = -\rho$. In what follows, it will be convenient to define the additional parameters\cite{copeland93,liddle94}
\begin{equation}
\eta \equiv {m_{\rm Pl}^2 \over 4 \pi} \left({H''\left(\phi\right) \over H\left(\phi\right)}\right)\label{eqdefofeta}
\end{equation}
and
\begin{equation}
\xi \equiv {m_{\rm Pl}^2 \over 4 \pi} \left({H'\left(\phi\right) H'''\left(\phi\right) \over H^2\left(\phi\right)}\right)^{1/2}.
\end{equation}
These are often referred to as ``slow-roll'' parameters, but the definition here is independent of the assumption of slow roll. 

The metric perturbations created during inflation are of two types: scalar, or curvature perturbations, which couple to the stress energy of matter in the universe and form the ``seeds'' for structure formation, and tensor, or gravitational wave perturbations, which do not couple to matter. Both scalar and tensor perturbations contribute to CMB anisotropy. Scalar fluctuations can also be interpreted as fluctuations in the density of the matter in the universe. The power spectrum of curvature perturbations is given by\cite{mukhanov92}
\begin{equation}
P_{\cal R}^{1/2}\left(k\right) = \sqrt{k^3 \over 2 \pi^2} \left|{u_k \over z}\right|
\end{equation}
where $k$ is a comoving wave number, and the mode function $u_k$ satisfies the differential equation\cite{mukhanov85,mukhanov88,stewart93}
\begin{equation}
{d^2 u_k \over d\tau^2} + \left(k^2 - {1 \over z} {d^2 z \over d\tau^2} \right) u_k = 0.\label{eqexactmodeequationa}
\end{equation}
The quantity $z$ is defined as
\begin{equation}
z \equiv 2 \sqrt{\pi} \left({a \dot\phi \over H}\right) = - m_{\rm Pl} a \sqrt{\epsilon},
\end{equation}
and
\begin{equation}
{1 \over z} {d^2 z \over d\tau^2} = 2 a^2 H^2 \left(1 + \epsilon - {3 \over 2} \eta + \epsilon^2 - 2 \epsilon \eta + {1 \over 2} \eta^2 + {1 \over 2} \xi^2\right).
\end{equation}
Solutions to the second-order differential equation for the mode $u_k$ in general contain two integration constants which can be taken to be phase and normalization. Normalization is fixed by the canonical quantization condition for the fluctuations, which in terms of the $u_k$ is a Wronskian condition
\begin{equation}
u_k^{*} {d u_k \over d\tau} - u_k {d u_k^{*} \over d\tau} = - i.\label{eqnormalizationcondition}
\end{equation}
Note that in the short wavelength limit, the equation of motion is just a wave equation. 
In a standard analysis, the phase is fixed at short wavelengths by selecting the so-called ``Bunch-Davies'' vacuum, which is equivalent to specifying that $u_k$ receive a contribution only from a negative-frequency component. 
\begin{equation}
u_k \propto e^{- i k \tau},\ k \rightarrow \infty.\label{largekbc}
\end{equation}
The solution in the long wavelength limit $k \rightarrow 0$ is just $u_k \propto z$. It is the second condition (\ref{largekbc}) that we relax in this analysis, allowing contributions from both positive- and negative-frequency components:
\begin{equation}
u_k \propto C_{\rm +} e^{- i k \tau} + C_{\rm -} e^{+ i k \tau},\ k \rightarrow \infty.\label{generallargekbc}
\end{equation}

The usual method of obtaining general solutions to the mode equation (\ref{eqexactmodeequationa}) is to solve for the quantity $\left(a H\right)$ as a function of the conformal time $\tau$. To do this, take the exact relation
\begin{equation}
d\tau = {d\left(a H\right) \over \left(a H\right)^2 \left(1 - \epsilon\right)}
\end{equation}
and integrate by parts:
\begin{eqnarray}
\tau =&& - {1 \over \left(a H\right) \left(1 - \epsilon\right)} + \int{{d\left(a H\right) \over \left(a H\right)} {d \over d\left(a H\right)} \left(1 \over 1 - \epsilon\right)}\cr
=&& - {1 \over \left(a H\right) \left(1 - \epsilon\right)} + \int{{2 \epsilon \left(\epsilon - \eta\right) \over \left(a H\right)^2 \left(1 - \epsilon\right)^3}\,d\left(a H\right)}.\label{eqtaubyparts}
\end{eqnarray}
In the limit of power-law inflation $\epsilon = \eta = {\rm const.}$ the second integral in Eq. (\ref{eqtaubyparts}) vanishes, and the conformal time is exactly
\begin {equation}
\tau = - {1 \over \left(a H\right) \left(1 - \epsilon\right)}.
\end{equation}
The mode equation (\ref{eqexactmodeequationa}) then becomes a Bessel equation, with the solution
\begin{equation}
u_k = {1 \over 2} \sqrt{- \pi \tau} \left[C_{+} H_\nu\left(- k \tau\right) + C_{-} H^*_\nu\left(- k \tau\right)\right],\label{eqhankelsolution}
\end{equation}
where $H_\nu$ is a Hankel function of the first kind, and
\begin{equation}
\nu =  {3 - \epsilon \over 2 \left(1 - \epsilon\right)}.
\end{equation}
Here we have normalized $u_k$ so that the Wronskian condition (\ref{eqnormalizationcondition}) is equivalent to the condition
\begin{equation}
\left|C_{+}\right|^2 - \left|C_{-}\right|^2 = 1\label{eqwronskiancp}
\end{equation}
on the constants $C_{\pm}$. Note that Eq. (\ref{eqwronskiancp}) is valid {\em in general}, not just in the short wavelength ($k \rightarrow \infty$) limit. Thus the choice of the Bunch-Davies vacuum at short wavelength is equivalent to setting $C_{-} = 0$ in the general solution (\ref{eqhankelsolution}).
The limit of de Sitter expansion is $\epsilon \rightarrow 0$, and this reduces to $\nu = 3/2$, which is the case of a scale invariant spectrum $P\left(k\right) \propto k$. Thus, de Sitter expansion can be considered to be a limiting case of power-law inflation. The so-called ``slow-roll expansion'' is an expansion in small parameters about the de Sitter limit. In cases where $\epsilon \neq \eta$, but both $\epsilon$ and $\eta$ are small, the conformal time is given by the (now approximate) relation
\begin {equation}
\tau \simeq - {1 \over \left(a H\right) \left(1 - \epsilon\right)} \simeq - {1 \over \left(a H\right)} \left(1 + \epsilon\right).
\end{equation}
Note that despite the formal similarity between this and the power-law case, slow roll involves distinct assumptions\cite{grivell96}: the slow-roll and power-law solutions are the same only in the de Sitter limit. Higher-order corrections can be obtained by continuing the integration by parts,
\begin{equation}
\tau = - {1 \over \left(a H\right) \left(1 - \epsilon\right)}\left[1 + {2 \epsilon \left(\epsilon - \eta\right) \over \left(1 - \epsilon\right)^2} + {\rm O}\left(\epsilon \eta^2\right) + \cdots\right].\label{eqseriesfortau}
\end{equation}
As long as this series converges, the conformal time is well defined as a series in slow-roll parameters. In the slow-roll approximation $\epsilon,\ \eta \ll 1$, it is consistent to take  $\epsilon$ and $\eta$ to be approximately constant, and the solutions are again Hankel functions of the form (\ref{eqhankelsolution}), with 
\begin{equation}
\nu = {3 \over 2} + 2 \epsilon - \eta\label{eqslowrollnu}
\end{equation}
to first order in the slow-roll parameters. 

Instead of expressing the mode equation (\ref{eqexactmodeequationa}) as a differential equation in the conformal time $\tau$, it is convenient to switch variables to the wavelength of the fluctuation mode relative to the horizon size\cite{kinney97},
\begin{equation}
y \equiv  \left({k \over a H}\right) \simeq \left({d_H \over \lambda}\right).
\end{equation}
Then
\begin{equation}
dy = - k {d \left(a H\right) \over \left(a H\right)^2}
= - k \left(1 - \epsilon\right) d\tau,\label{eqdifferentialy}
\end{equation}
and the mode equation (\ref{eqexactmodeequationa}) can be expressed exactly as
\begin{equation}
y^2 \left(1 - \epsilon\right)^2 {d^2 u_k \over d y^2} + 2 y \epsilon \left(\epsilon - \eta\right) {d u_k \over d y} + \left[y^2 - F\left(\epsilon, \eta, \xi\right)\right] u_k = 0,\label{eqexactmodeequationya}
\end{equation}
where 
\begin{equation}
F\left(\epsilon,\eta,\xi\right) \equiv 2 \left(1 + \epsilon - {3 \over 2} \eta + \epsilon^2 - 2 \epsilon \eta + {1 \over 2} \eta^2 + {1 \over 2} \xi^2\right).
\end{equation}
The case of tensor fluctuations is exactly analogous. The tensor fluctuation amplitude is defined as:
\begin{equation}
P_{\cal T}^{1/2}\left(k\right) = \sqrt{k^3 \over 2 \pi^2} \left|{v_k \over m_{\rm Pl} a}\right|,
\end{equation}
where the mode function $v_k$ obeys the equation of motion:
\begin{equation}
{d^2 v_k \over d\tau^2} + \left(k^2 - {1 \over a} {d^2 a \over d\tau^2} \right) v_k = 0.\label{eqtenexactmodeequationa}
\end{equation}
This can be written in a form similar to Eq. (\ref{eqexactmodeequationya}),
\begin{equation}
y^2 \left(1 - \epsilon\right)^2 {d^2 v_k \over d y^2} + 2 y \epsilon \left(\epsilon - \eta\right) {d v_k \over d y} + \left[y^2 - \left(2 - \epsilon\right)\right] v_k = 0.\label{eqtenexactmodeequationya}
\end{equation}
This equation is also exact. We can then take Eqs. (\ref{eqexactmodeequationy}) and (\ref{eqtenexactmodeequationy}) as the starting point for an analysis of the effect of the choice of vacuum at short wavelengths.
\end{document}